\documentclass{appolb}
\usepackage{epsfig}
\begin{document}
\title{Transport properties of hot gluonic matter%
\thanks{Contribution to 
{\it Excited QCD 2012}, Peniche, Portugal, May 6-12, 2012.}%
}
\author{M. Bluhm
\address{SUBATECH, UMR 6457, Universit\'{e} de Nantes,
Ecole des Mines de Nantes, IN2P3/CNRS. 4 rue Alfred Kastler,
F-44307 Nantes cedex 3, France}
\address{CERN, Physics Department, Theory Devision, 
CH-1211 Geneva 23, Switzerland}
\and
M. Nahrgang
\address{SUBATECH, UMR 6457, Universit\'{e} de Nantes,
Ecole des Mines de Nantes, IN2P3/CNRS. 4 rue Alfred Kastler,
F-44307 Nantes cedex 3, France}
\address{Frankfurt Institute for Advanced Studies (FIAS), Ruth-Moufang-Str. 1, 
60438 Frankfurt am Main, Germany}
}
\maketitle
\begin{abstract}
\noindent
We discuss the temperature dependence of the scaled jet quenching parameter of 
hot gluonic matter within a quasiparticle approach. A pronounced maximum in the vicinity of the
transition temperature is observed, where the ratio of the scaled jet quenching parameter and the inverse specific shear viscosity increases above typical values for weakly coupled systems. 
\end{abstract}

\section{Introduction \label{sec:1}}

Understanding the nature of QCD matter at high temperatures and densities means understanding the 
relevant degrees of freedom and their interaction. Immense efforts have been invested both in 
theory and in experiment in order to study the transport properties of strongly interacting matter, 
which can be described by certain transport coefficients. They determine the evolution and relaxation 
of conserved charge densities within dynamical systems. Shear ($\eta$) and bulk ($\zeta$) viscosities, 
for example, describe the hydrodynamic response of a medium to energy and momentum density 
fluctuations. A firm knowledge of these transport coefficients would have a wide range of implications 
in cosmology, astrophysics and in nuclear physics. 

Measurements at RHIC and LHC revealed that the matter produced in relativistic heavy-ion collisions 
behaves like an almost perfect fluid, cf.~e.g.~the reviews in~\cite{Teaney09,Muller12}. Indeed, the 
measured flow pattern is best described by assuming only small dissipative effects with a specific 
shear viscosity $\eta/s$ ($s$ is the entropy density in the system) of at most a few times the 
conjectured lower (KSS-)bound~\cite{Kovtun03} $(\eta/s)_{KSS}=1/(4\pi)$ which is based on holographic 
principles. 

Another outstanding observation made in central collisions of relativistic heavy nuclei is 
that final state hadrons are strongly suppressed at large transverse 
momenta~\cite{Adcox01,Adler02,Aamodt11,Chatrchyan12,Steinberg11}. This phenomenon 
of jet quenching indicates that the produced matter is opaque, i.e.~that coloured 
charges suffer from a sizeable energy loss while traversing the 
medium~\cite{Majumder11}. An important transport coefficient in the context of energy loss 
is the jet quenching parameter~\cite{Baier97,Arnold08} 
$\hat{q}$ defined as the average squared transverse momentum transfer 
to a highly energetic particle per unit length due to multiple elastic scatterings with medium constituents. 
The parameter $\hat{q}$, thus, governs the transverse momentum broadening of 
propagating colour charges per unit length~\cite{Baier97}. 
The probability for the formation of gluon bremsstrahlung depends sensitively on the 
efficiency by which a bremsstrahlung gluon decoheres from its emitter. As random 
scatterings with medium constituents accelerate this process, also the radiative energy 
loss of partons in the multiple scattering regime is governed by $\hat{q}$. 

An intriguing question is whether different transport coefficients are fundamentally 
related with each other. 
In this work, we study the possible connection between 
$\eta$ and $\hat{q}$ within a quasiparticle approach, focussing on the case of pure hot, deconfined gluonic matter. In 
particular, the influence of a temperature dependent quasiparticle dispersion 
relation is investigated. 

\section{Relating shear viscosity and jet quenching parameter \label{sec:2}}

In~\cite{Majumder07} it was demonstrated that $\eta$ and $\hat{q}$ are formally 
connected, when one assumes (i) that the strongly interacting medium is effectively 
describable within a partonic quasiparticle framework and (ii) that the interaction 
between an energetic parton and the partonic quasiparticles is of the same structure 
and strength as the interaction among the medium constituents. In this case, one finds 
the relation~\cite{Majumder07} 
\begin{equation}
 \label{equ:relagen}
 \frac{\hat{q}}{T^3} \simeq 
 \frac{\rho}{12\,s}\frac{\langle p\rangle \langle\hat{s}\rangle}{T^3}
 \left(\frac{\eta}{s}\right)^{-1} ,
\end{equation}
where $\rho$ is the quasiparticle density in the thermal medium with temperature 
$T$, $\langle p\rangle$ the average quasiparticle momentum and $\langle\hat{s}\rangle$ 
the average center-of-mass energy squared in soft, elastic quasiparticle scatterings. 

When, in particular, a thermal ensemble of free massless bosons is considered, 
Eq.~(\ref{equ:relagen}) becomes 
\begin{equation}
 \label{equ:freemasslessbosons}
 \frac{\hat{q}}{T^3} \approx 1.25 \left(\frac{\eta}{s}\right)^{-1} .
\end{equation}
According to~\cite{Majumder07}, this relation is generally true for any weakly coupled 
partonic quasiparticle system, while for a strongly coupled system 
$\hat{q}/T^3\gg(\eta/s)^{-1}$ is conjectured. In the following, we will investigate 
the relation Eq.~(\ref{equ:relagen}) within a phenomenological quasiparticle model 
(QPM)~\cite{Bluhm07-1,Bluhm07-2}, in which gluonic quasiparticles obey a temperature 
dependent dispersion relation. 

\section{Viscosity coefficients of hot gluonic matter within a quasiparticle picture 
\label{sec:3}}

In quantum field theories, transport coefficients such as the shear or bulk viscosity 
can be calculated within the framework of linear response theory from 
Kubo-relations~\cite{Kubo57,Hosoya84}. 
Likewise, kinetic theory, e.g.~in the form of a linearized Boltzmann equation, 
can be applied as a rigorous tool for evaluating the transport properties of systems 
with weak interaction strength~\cite{deGroot}. Here, the underlying assumption 
is that the dynamics of the system is describable in terms of weakly interacting 
quasiparticle excitations. Such a 
picture leads to a lower bound in $\eta/s$ as consequence of the uncertainty 
principle, cf.~\cite{Danielewicz85}. 

By extrapolating perturbative QCD results for $\eta/s$~\cite{Arnold03} towards the confinement-deconfinement transition temperature $T_c$~\cite{Csernai06} one cannot explain the small values deduced from experiment.
This, however, 
does not necessarily imply that any picture based on quasiparticle excitations is 
inadequate for describing the properties of the produced matter. Within the 
QPM~\cite{Bluhm07-1,Bluhm07-2}, based on an effective kinetic theory approach of 
Boltzmann-Vlasov type, the shear and bulk viscosities have been determined 
in~\cite{BluhmVisc} for the pure gluon plasma in relaxation time approximation and their 
ratio was studied in~\cite{BluhmPLB}. In this approach, the medium is viewed as being 
composed of quasigluons with the same quantum numbers and obeying the same quantum 
statistics as partonic gluons. They follow, however, a medium-modified dispersion relation featuring 
a temperature-dependent gluon mass which accommodates non-perturbative effects in the 
vicinity of $T_c$, cf.~\cite{Bluhm07-2}. 

A fairly good quantitative agreement with the available lattice QCD results~\cite{Nakamura05,Meyer07-08} was 
obtained~\cite{BluhmVisc}. Both, specific shear and bulk viscosities, were found to exhibit 
a pronounced behaviour. At high $T$, $\eta/s$ is large and shows the parametric dependencies known from perturbative QCD~\cite{Arnold03}, while it has a minimum near $T_c$.
In contrast, $\zeta/s$ is found to become large near $T_c$, similar to~\cite{Karsch08}, while it vanishes logarithmically with increasing 
temperature.

\section{Jet quenching parameter \label{sec:4}}

\begin{figure}[t]
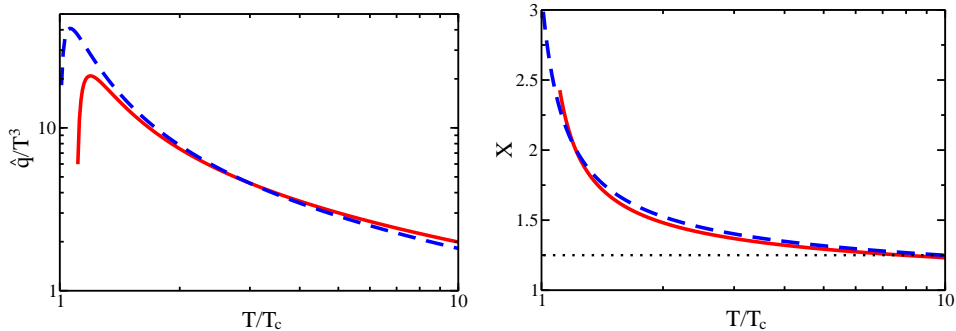

\centering
\includegraphics[width=0.48\textwidth]{qhat3_1.eps} 
\hspace{1mm}
\includegraphics[width=0.48\textwidth]{qhat4_1.eps} 
\caption[]{\label{fig:1} Left: Scaled jet quenching parameter $\hat{q}/T^3$ as a function 
of $T/T_c$. Right: Ratio $X$, cf.~text for details.}
\end{figure} 

Via Eq.~(\ref{equ:relagen}), the scaled jet quenching parameter $\hat{q}/T^3$ can be obtained 
from $\eta/s$ calculated within the QPM for pure gluodynamics. 
The result is shown 
in Fig.~\ref{fig:1} (left panel). The pronounced maximum in $\hat{q}/T^3$ is a consequence 
of the minimum in $\eta/s$ in the vicinity of $T_c$, cf.~\cite{BluhmVisc}. The solid curve is 
obtained for the QPM and relaxation time parametrizations used in~\cite{BluhmVisc}, whereas 
the dashed curve is the result of the corresponding parametrizations advocated 
in~\cite{Khvorostukhin11}. We note that for the latter, $\eta/s$ is quantitatively 
comparable with the results presented in~\cite{BluhmVisc}, however, its minimum value is 
slightly below the KSS-bound. 

As evident from the left panel of Fig.~\ref{fig:1}, the scaled jet quenching parameter 
is significantly enhanced just above $T_c$ as compared to the high-temperature region, 
while for $T\to T_c^+$ it drastically drops down again. This implies that the energy 
loss of energetic partons depends sensitively on the temperature of the surrounding matter. 
Such a behaviour could have a strong impact on related observables. For example, 
in~\cite{Scardina11} it was argued that for a simultaneous description of both nuclear 
modification factor and elliptic flow of hadrons an energy loss mainly occuring around 
$T_c$ is crucial. Similar conclusions were drawn in~\cite{Liao09}. 

As proposed in~\cite{Majumder07}, the ratio $X=(\hat{q}/T^3)/(\eta/s)^{-1}$ could indicate if
 the considered medium is a weakly or a strongly coupled system. 
In line with the non-trivial temperature dependence in $\eta/s$ and $\hat{q}/T^3$ in the QPM, 
the ratio $X$ is also a function of $T$. This is shown in Fig.~\ref{fig:1} (right panel), 
where we compare $X$ obtained for the two different parametrization sets 
from~\cite{BluhmVisc} (solid curve) and from~\cite{Khvorostukhin11} (dashed curve) with the 
constant value $X\approx 1.25$ for a gas of free massless bosons (dotted horizontal line), 
cf.~Eq.~(\ref{equ:freemasslessbosons}). At high temperatures, the ratio $X$ in the QPM 
approach becomes as small as $1.25$ indicating that the described system is indeed weakly 
coupled. However, as $T\to T_c^+$, $X$ increases significantly as a consequence of the 
medium-modified gluon dispersion relation. For strong coupling, based on analytic 
results for $\eta/s$~\cite{Buchel05} and $\hat{q}$~\cite{Liu06} obtained in $\mathcal{N}=4$ 
supersymmetric Yang-Mills theory in the limit of large 't Hooft coupling $\lambda$, one 
finds for the ratio $X=\Gamma(3/4)\sqrt{\lambda\pi}/(4\,\Gamma(5/4))$, which is large 
for large $\lambda$. For a reasonably chosen value $\lambda=6\pi$ cf.~\cite{Liu06}, however, 
one finds $X\approx 2.602$ which can also be obtained within the QPM, cf. Fig.~\ref{fig:1} 
(right panel). 

\section{Conclusions \label{sec:5}}

We studied the temperature dependence of 
the jet quenching parameter $\hat{q}$ of hot, deconfined gluonic matter within a phenomenological quasiparticle approach. For this purpose, 
we made use of a general relation between $\hat{q}$ and the shear viscosity $\eta$ valid for 
any weakly coupled partonic quasiparticle system~\cite{Majumder07}. From $\eta$, as previously 
determined in the QPM~\cite{BluhmVisc}, $\hat{q}/T^3$ is found to show a pronounced 
maximum close to $T_c$, while falling off steeply with increasing $T$. This implies 
that the energy loss of energetic partons traversing the coloured medium exhibits a significant 
temperature dependence. The relation between $\hat{q}$ and $\eta$ expressed by the ratio $X$ shows a non-trivial 
behaviour with $T$ due to the medium-modified gluon dispersion relation. At large $T$, $X$ is as small as typically assumed for weakly coupled systems~\cite{Majumder07}, while as $T\to T_c^+$ it becomes quantitatively comparable with reasonable estimates for strongly coupled systems~\cite{Majumder07, Liu06}.

\end{document}